\documentclass[prl,twocolumn,times,showpacs,showkeys,superscriptaddress,citeautoscript]{revtex4-1}

\usepackage{amsmath,amssymb}
\usepackage[colorlinks=true,citecolor=blue,linkcolor=blue]{hyperref}
\usepackage{graphicx}
\usepackage{physics}
\usepackage{tabularx}

\usepackage{array}
\newcolumntype{R}[1]{>{\raggedleft\let\newline\\\arraybackslash\hspace{0pt}}m{#1}}

\usepackage[inline]{enumitem}

\newcommand*{\setR}{\ensuremath{\mathbb{R}}}

 \usepackage[textsize=footnotesize]{todonotes}
 \RequirePackage{framed}

 \usepackage{soul}
\usepackage{microtype}

\begin{document}

\title{Conformal dimensions via large charge expansion}

\author{Debasish Banerjee}
\email{debasish.banerjee@desy.de}
\affiliation{\textsc{nic}, \textsc{desy}, Platanenallee 6, \textsc{d}-15738 Zeuthen, Germany}
\author{Shailesh Chandrasekharan}
\email{sch@phy.duke.edu}
\affiliation{Department of Physics, Duke University, Durham, North Carolina 27708, \textsc{usa}}
\author{Domenico Orlando}
\email{dorlando@itp.unibe.ch}
\affiliation{Albert Einstein Center for Fundamental Physics, Institute for Theoretical Physics, University of Bern,
Sidlerstrasse 5, \textsc{ch}-3012,Bern, Switzerland}
\affiliation{INFN, Sezione di Torino, Via Pietro Giuria 1, 10125 Torino, Italy}

\begin{abstract}
  We construct an efficient Monte Carlo algorithm that overcomes the severe signal-to-noise ratio problems and helps us to accurately compute the conformal dimensions of large-$Q$ fields at the Wilson--Fisher fixed point in the $O(2)$ universality class. Using it we verify a recent proposal that conformal dimensions of strongly coupled conformal field theories with a global $U(1)$ charge can be obtained via a series expansion in the inverse charge $1/Q$. We find that the conformal dimensions of the lowest operator with a fixed charge $Q$ are almost entirely determined by the first few terms in the series.
\end{abstract}
\preprint{}
\maketitle

Conformal field theories (CFTs) occupy a central place in our understanding of modern physics. They describe critical phenomena in condensed matter physics and statistical models~\cite{Cardy:2008jc,Pelissetto2002}, quantum gravity via the AdS/CFT correspondence~\cite{Maldacena:1997re} and can be found at fixed points of renormalization group flows~\cite{Zamolodchikov:1986gt,Polchinski:1987dy,Nakayama20151,Rychkov:2016iqz}. They are uniquely described by a set of dimensionless numbers (the CFT data), \emph{i.e.} conformal dimensions and OPE coefficients associated with the primary fields of the theory. Since they are typically strongly coupled and lack a characteristic scale, it is often difficult to compute the conformal dimensions analytically. Still, a number of sophisticated techniques have been developed to deal with this challenge, both perturbatively (\emph{e.g.} \(4-\epsilon\) expansion, fixed-dimension expansion, large-\(N\); see~\cite{Pelissetto:2000ek} for a review) and non-perturbatively (\emph{e.g.} bootstrap~\cite{Rattazzi:2008pe}). In some cases Monte Carlo techniques offer a reliable numerical approach for computing the conformal dimensions~\cite{Campostrini:2000iw,Campostrini:2002ky}.

Energies of low-lying states also capture universal features of a $2+1$ dimensional CFT when the theory is studied on a space-time manifold \(\setR \times \Sigma\) (see~\cite{PhysRevLett.117.210401,Whitsitt:2017ocl} for some recent work in this direction).
For example conformal dimensions \(D\) of operators on \(\setR^3\) are related to the energies \(E_{S^2}\) of states living on a two-sphere of radius \(r_0\) through the relation \(D = r_0 E_{S^2}\)~\cite{Cardy:1984rp,Cardy:1985lth}.
This relation, known as the state-operator correspondence, is a consequence of the fact that \(\setR^3\) is conformally equivalent to \(\setR \times S^2(r_0)\).
Recently, such a connection has been used in CFTs with global \(U(1)\) charges to show that the conformal dimension $D(Q)$ of the lowest operator with fixed \(U(1)\) charge \(Q\) can be expanded in inverse powers of the charge density on a unit sphere $Q/4\pi$~\cite{Hellerman:2015nra,Alvarez-Gaume:2016vff} (see also~\cite{Monin:2016jmo,Loukas:2016ckj,Hellerman:2017efx,Hellerman:2017veg,Loukas:2017lof} for related work):
\begin{equation}
D(Q) = \sqrt{\frac{Q^3}{4\pi}}
\left(c_{\frac{3}{2}} + c_{\frac{1}{2}} \left(\frac{4\pi}{Q}\right) %
  + \dots \right) + c_0 + \order{\frac{1}{Q}}
\label{eq:conformal-dimensions}
\end{equation}
where $c_0 \approx -0.094$~\cite{Monin:2016bwf} and the other coefficients only depend on the universality class. While a simple dimensional analysis allows one to predict the leading large-$Q$ behavior, it is a priori unclear if a power series can capture the sub-leading corrections.
The recent work argues that by separating the theory into sectors of fixed charge \(Q\) one can construct an effective field theory (EFT) in each sector, which can be used to compute the energies as a power series in $1/Q$. Through the state-operator correspondence one can then obtain the series expansion of the conformal dimension $D(Q)$ and relate directly the coefficients in Eq.~\eqref{eq:conformal-dimensions} to the coefficients in the energy expansion. 

\begin{figure}[!htb]
 \includegraphics[width=0.45\textwidth]{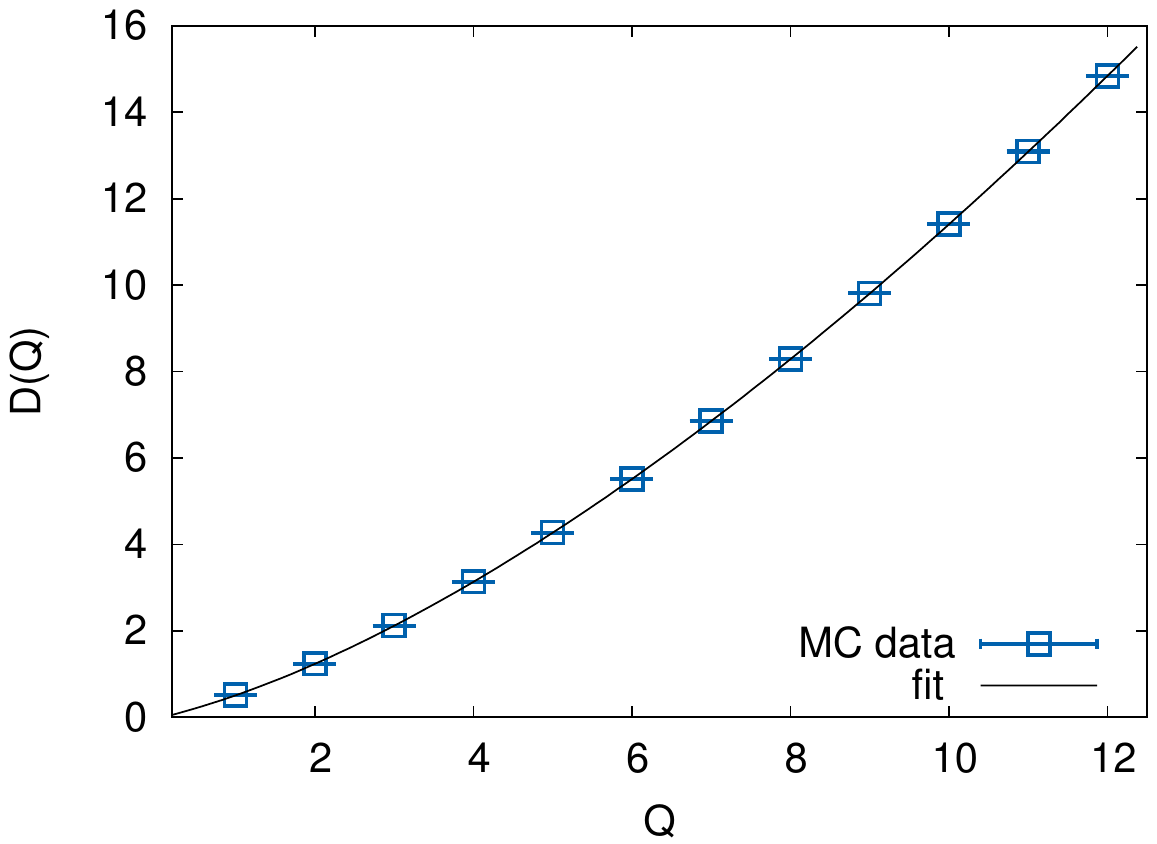}
 \caption{\label{fig:confD} Plot of the values of $D(Q)$ extracted from our Monte Carlo calculations at the $O(2)$ Wilson--Fisher (WF) fixed point, along with the plot of Eq.~\eqref{eq:conformal-dimensions} (solid line) with our estimated values $c_{\frac{3}{2}} = 1.195$ and $c_{\frac{1}{2}}=0.075$ and previously computed value $c_0=-0.094$. It is surprising that these three leading coefficients in  Eq.~\eqref{eq:conformal-dimensions} can predict the conformal dimensions for all $Q\geq 1$ very well.}
 \end{figure}

In this work we make significant progress in establishing that Eq.~(\ref{eq:conformal-dimensions}) is an excellent description of the WF fixed point in the $O(2)$ universality class. In order to achieve this we overcome severe signal-to-noise ratio problems in Monte Carlo methods that usually hinder calculations of $D(Q)$ for large values of $Q$. Our new approach allows us to determine the corresponding universal coefficients $c_{\frac{3}{2}}=1.195(10)$, $c_{\frac{1}{2}}=0.075(10)$ for the first time. In Fig.~\ref{fig:confD} we demonstrate that the measured values of $D(Q)$ using our Monte Carlo approach are excellently described by Eq.~(\ref{eq:conformal-dimensions}). Surprisingly, we find that the large-$Q$ expansion with the first three coefficients matches the conformal dimensions even at $Q = 1$ within a few percent. Thus, our work demonstrates that at least in some class of models, the large Q expansion is similar to the epsilon expansion in the fact that the strongly-coupled conformal fixed point can be described within a simple perturbative framework. The coefficients of the expansion could still be difficult to compute analytically, but perhaps bootstrap techniques could be developed for it~\cite{Jafferis:2017zna}.

To understand the origin of Eq.~(\ref{eq:conformal-dimensions}), consider the conformal field theory describing the Wilson--Fisher fixed point in the three-dimensional $O(2)$ universality class. In a fixed charge sector $Q$, the charge density introduces the mass scale $\sqrt{Q/V}$ in the theory and hence for momentum scales $p \ll \sqrt{Q/V}$ the physics is described by a Goldstone field \(\chi\) that is controlled by an approximately scale-invariant Lagrangian~\cite{Hellerman:2015nra,Monin:2016jmo} (see also~\cite{Son:2005rv} for a related approach to effective descriptions of non-relativistic CFTs):
\begin{equation}
  \label{eq:effective-Lagrangian}
  L = \frac{k_{\frac{3}{2}}}{27} \pqty{ \partial_\mu \chi \partial^\mu \chi}^{\frac{3}{2}} + \frac{k_{\frac{1}{2}} R}{3} \pqty{ \partial_\mu \chi \partial^\mu \chi}^{\frac{1}{2}} + \dots
\end{equation}
where \(R\) is the scalar curvature of the manifold  \(\setR \times \Sigma\). Thus, we learn that in the large-$Q$ limit only the two parameters \(k_{\frac{3}{2}}\) and \(k_{\frac{1}{2}}\) that appear in Eq.~(\ref{eq:effective-Lagrangian}) play an important role and all other terms are suppressed~\cite{Alvarez-Gaume:2016vff,Loukas:2016ckj}. Since the charge is non-zero, the action is only meaningful away from \(\chi = 0\) and is to be expanded around the fixed-charge homogeneous classical solution \(\chi = \mu t\). 
Using the effective quantum Hamiltonian arising from the effective Lagrangian Eq.~\eqref{eq:effective-Lagrangian}, one can show that the total energy of the system is given by
\begin{equation}
  \label{eq:classical-energy}
  E_{\Sigma}(Q) = \sqrt{\frac{Q^3}{V}}\left( c_{\frac{3}{2}}+ 
    c_{\frac{1}{2}} \left(\frac{RV}{2Q}\right) + \dots \right) + q_\Sigma + \order{\frac{1}{Q}} ,
\end{equation}
where the first two terms are related to the couplings in the effective Lagrangian Eq.~\eqref{eq:effective-Lagrangian} through the relations \(k_{\frac{3}{2}} = 4 /c_{\frac{3}{2}}^2 \) and \(k_{\frac{1}{2}} = - c_{\frac{1}{2}}/c_{\frac{3}{2}}\). The higher order terms in the expansion are related to higher dimensional operators in Eq.~(\eqref{eq:effective-Lagrangian}) and quantum corrections. The last term $q_\Sigma$ arises due to quantum fluctuations that can be computed exactly for simple manifolds. For the sphere ($R=2/r_0^2$) one finds $q_{S^2} = c_0/r_0$ where $c_0 \approx -0.094$ \cite{Monin:2016bwf}, while for the torus ($R = 0$) it is $q_{T^2} = c_0/L$ with $c_0 \approx -0.508$~\cite{Bordag:2009zzd}. 

By choosing $\Sigma = S^2$ and using the state operator correspondence one can now easily derive Eq.~\eqref{eq:conformal-dimensions}. It is interesting to note that the coefficients $c_{\frac{3}{2}}$, $c_{\frac{1}{2}}$ in Eqs.~\eqref{eq:conformal-dimensions},\eqref{eq:classical-energy} are related to the low-energy constants $k_{\frac{3}{2}}$ and $k_{\frac{1}{2}}$ of the effective Lagrangian in Eq.~\eqref{eq:effective-Lagrangian}. Indeed, these low-energy constants are independent of the manifold chosen and depend only on the CFT. Assuming the manifold is the torus we predict that
\begin{equation}
  \lim_{Q \to \infty }\frac{D(Q)}{E_{T^2}(Q) L} = \frac{1}{2\sqrt{\pi}}.
 \label{eq:dimensions-vs-energy}
\end{equation}
Note that every term in the energy expansion is a dimensionless function of three variables: a coefficient in the $D(Q)$ expansion, a geometrical term from the manifold, and a power of $V/Q$.

The motivation of our current work is to compute $D(Q)$ and $E_{\Sigma}(Q)$ in the classical $O(2)$ sigma model on a torus and verify the expansions in Eq.~\eqref{eq:conformal-dimensions} and~\eqref{eq:classical-energy} and the connections between them. We accomplish this by regularizing the classical $O(2)$ sigma model on a cubic lattice with lattice spacing $a$ and use Monte Carlo methods to perform the calculations. The model is defined by phases, $\exp(i \theta_\mathrm{x})$ on each three-dimensional lattice site $\mathrm{x} = (x_1 a,x_2 a,x_3 a)$ and the nearest neighbor action 
\begin{equation}
  S = -\beta \sum_{\mathrm{x},\alpha} \cos( \theta_{\mathrm{x}} - \theta_{\mathrm{x}+\hat{\alpha} a}).
 \label{eq:act}
\end{equation}
Here $\hat{\alpha} a$ denotes the three unit lattice vectors, and $\beta$ is the coupling of the model. The physics of the Wilson--Fisher fixed point can be studied by tuning the coupling to its critical value ($\beta_c = 0.4541652$~\cite{Hasenbusch:1999cc,Campostrini:2006ms,Deng:2005dh}), where a second-order phase transition separates the symmetric phase ($\beta < \beta_c$) and the spontaneously broken superfluid phase ($\beta > \beta_c$). Universality implies that details our specific model should be irrelevant in the limit $a \rightarrow 0$ which is naturally reached by studying large lattices at $\beta_c$.

Configurations that contribute to the partition function of the lattice model at the critical point can be efficiently generated by both the Wolff cluster algorithm~\cite{Wolff:1988uh} and the worm algorithm based on the worldline representation~\cite{PhysRevD.81.125007}. However, in order to compute the conformal dimension $D(Q)$ in \(\setR^3\) we need to compute the two-point correlation function $C_Q (r)$ of charge $Q$ fields on a large lattice of size $L$, which is expected to decay as a power law for large separations $r \ll L$ at the critical point:
 \begin{equation}    
 C_Q (r) = \ev{ \exp( i Q \theta_{\mathrm{r}})  \exp(-i Q \theta_{\mathrm{0}}) }
\sim \frac{a(Q)}{|\mathbf{r}|^{2 \cdot D(Q)}}.
 \label{powL}
\end{equation}
Fitting the data to this form we can in principle extract $D(Q)$ and thus verify Eq.~\eqref{eq:conformal-dimensions}. Note that for $Q=1$, it reduces to the standard 2-point correlation function, which is used to extract the critical exponent $\eta$ through the relation $C_1 (r) = G(r) \propto 1/r^{d-2+\eta}$. For $Q = 2, 3, 4$, the corresponding conformal dimensions have also been computed earlier using different methods, and the results are summarized in Table \ref{tab:dq-check}.
Unfortunately, calculations of $D(Q)$ for higher values of $Q$ do not exist and hence the relation~\eqref{eq:conformal-dimensions} remains unconfirmed.
\begin{table}[htb]
   \centering
   \begin{tabular}{rR{5em}R{5em}R{5em}R{5em}}
     \toprule
     \(Q\) & \(\epsilon^5\) & \(\lambda^6\) & MC        & bootstrap \\
     \colrule
     1     & 0.518(1)       & -             & 0.5190(1) & 0.5190(1) \\
     2     & 1.234(3)       & 1.23(2)       & 1.236(1)  & 1.236(3)  \\
     3     & 2.10(1)        & 2.10(1)       & 2.108(2)  & -         \\
     4     & 3.114(4)       & 3.103(8)      & 3.108(6)  & -         \\
     \botrule
   \end{tabular}
\caption{Conformal dimensions $D(Q)$ obtained previously by other methods for $Q \leq 4$: Field theory results in \(4 - \epsilon\) dimensions at five loops are in column two, six loop results at \(d = 3\) are in column three (\cite{Calabrese:2002bm} for \(Q = 2\), in~\cite{DePrato:2003yd} for \(Q = 3\) and in~\cite{Carmona:1999rm} for \(Q = 4\)), previous MC results are in column four~\cite{Hasenbusch:2011xxx}, and bootstrap results are in column five~\cite{Kos:2015mba}.}
   \label{tab:dq-check}
 \end{table}

\begin{figure}[!htb]
  \includegraphics[scale=0.65]{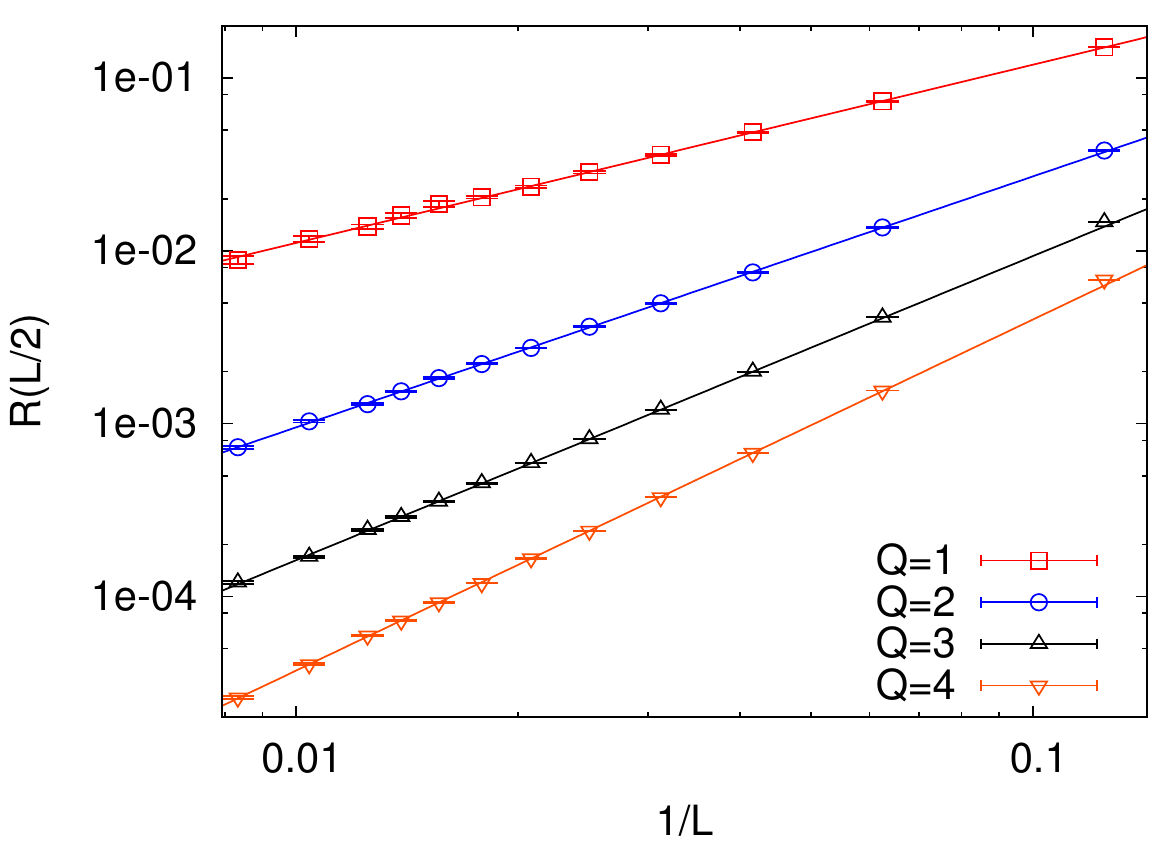}
  \includegraphics[scale=0.65]{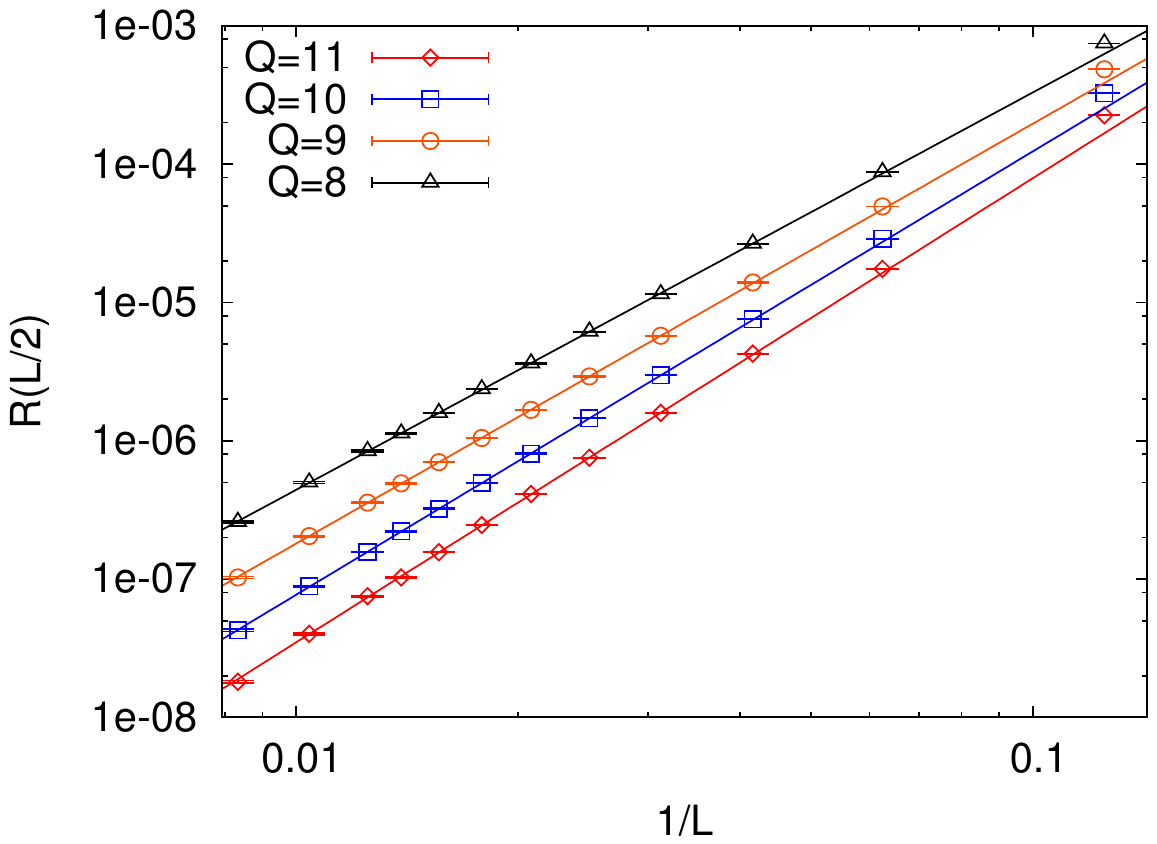}
  \caption{The figures (top and bottom) show the quantity R(L/2) for different lattice sizes $L/a = 8, \ldots, 120$
 and different $Q$ values. The straight line on a log-log plot is indicative of the power law behavior, and the slope gives the difference of the conformal dimensions $2 \Delta (Q)$. Note that there is no visible signal-to-noise problem in these correlators.}
  \label{chQ}
\end{figure}

It is difficult to measure $D(Q)$ for large values of $Q$ through a Monte Carlo method due to severe signal-to-noise ratio problems in the Monte Carlo calculations. With the Wolff cluster algorithm it is difficult to average numbers of order one to compute a small value of $C_Q(r)$ at large separations. In contrast, in the worm algorithm, it is difficult to correctly build the worldline configurations that contribute to the correlation function in the presence of charged sources $Q$ and $-Q$ separated by a large distance. In this case the severe signal-to-noise ratio problem emerges as an overlap problem between the vacuum ensemble and the one containing the sources. In order to overcome this problem we have designed an algorithm to efficiently compute the ratio
\begin{equation}
R(L/2) = \frac{C_{Q}(r=L/2)}{C_{Q-1}(r=L/2)}  
\end{equation}
on cubic lattices of side $L$ for $8 \leq L/a \leq 120$ at the critical point $\beta_c$ (the details of our algorithm can be found in the supplementary material). Using $R(L/2)$ it is easy to extract the difference $\Delta(Q) = D(Q)-D(Q-1)$ using the relation $R(L) \sim 1/L^{2{\Delta(Q)}}$. The accuracy with which we are able to compute the ratio $R(L/2)$ for various values of $Q$ can be seen in  Fig.~\ref{chQ}. Once the differences $\Delta(Q)$ are known, we can also extract $D(Q)$ by setting $D(Q=0) = 0$. Our estimates of both $\Delta(Q)$ and $D(Q)$ using Monte Carlo calculations, are given in Table~\ref{tab:Qresults}.  It is easy to verify that our results match quite well with previous results in Table~\ref{tab:dq-check} when $Q < 4$.
\begin{table}[!htb]
  \begin{tabular}{ccc|ccc}
    \toprule
    $Q$    & $\Delta(Q)$  & $D(Q)$   & $Q$    & $\Delta(Q)$  & $D(Q)$               \\
    \colrule
    1      & 0.516(3)       & 0.516(3)             &
    7      & 1.332(5)       & 6.841(8)             \\
    2      & 0.722(4)       & 1.238(5)             &
    8      & 1.437(4)       & 8.278(9)             \\
    3      & 0.878(4)       & 2.116(6)             &
    9      & 1.518(2)       & 9.796(9)             \\
    4      & 1.012(2)       & 3.128(6)             &
    10     & 1.603(2)       & 11.399(10)           \\
    5      & 1.137(2)       & 4.265(6)             &
    11     & 1.678(5)       & 13.077(11)           \\
    6      & 1.243(3)       & 5.509(7)             &
    12     & 1.748(5)       & 14.825(12)           \\
    \botrule
  \end{tabular}
  \caption{\label{tab:Qresults} Results for the conformal dimensions $\Delta(Q)$ and $D(Q)$ defined through~\eqref{powL}. Fit systematics are discussed in the Supplementary Material.  While our results for $Q < 4$ are in good agreement with previous results as seen in Table~\ref{tab:dq-check}, there is a slight deviation for $Q=4$.}
\end{table}

We can now verify if the conformal dimensions in Table~\ref{tab:Qresults} are consistent with Eq.~\eqref{eq:conformal-dimensions}. For this purpose we perform a combined fit of our data for the difference $\Delta(Q)$ and $D(Q)$ assuming that  $c_{\frac{3}{2}},c_{\frac{1}{2}},c_{-\frac{1}{2}}$ are non-zero and $c_0=-0.094$ as expected. Taking into account various systematic errors from fitting procedures we estimate $c_{\frac{3}{2}} = 1.195(10)$, $c_{\frac{1}{2}} = 0.075(10)$ and $c_{-\frac{1}{2}} = 0.0002(5)$. The raw data for $\Delta(Q)$ are shown in Fig.~\ref{fig:dqfits}, and further technical details are discussed in the Supplementary Material. We also show a comparison with the prediction obtained by just keeping the first three leading terms of the expansion in Eq.~\eqref{eq:conformal-dimensions}. As the figure shows, this prediction works even at small values of $Q$ but is off only by a few percent at $Q=1$.

 \begin{figure}[!htb]
 \includegraphics[width=0.45\textwidth]{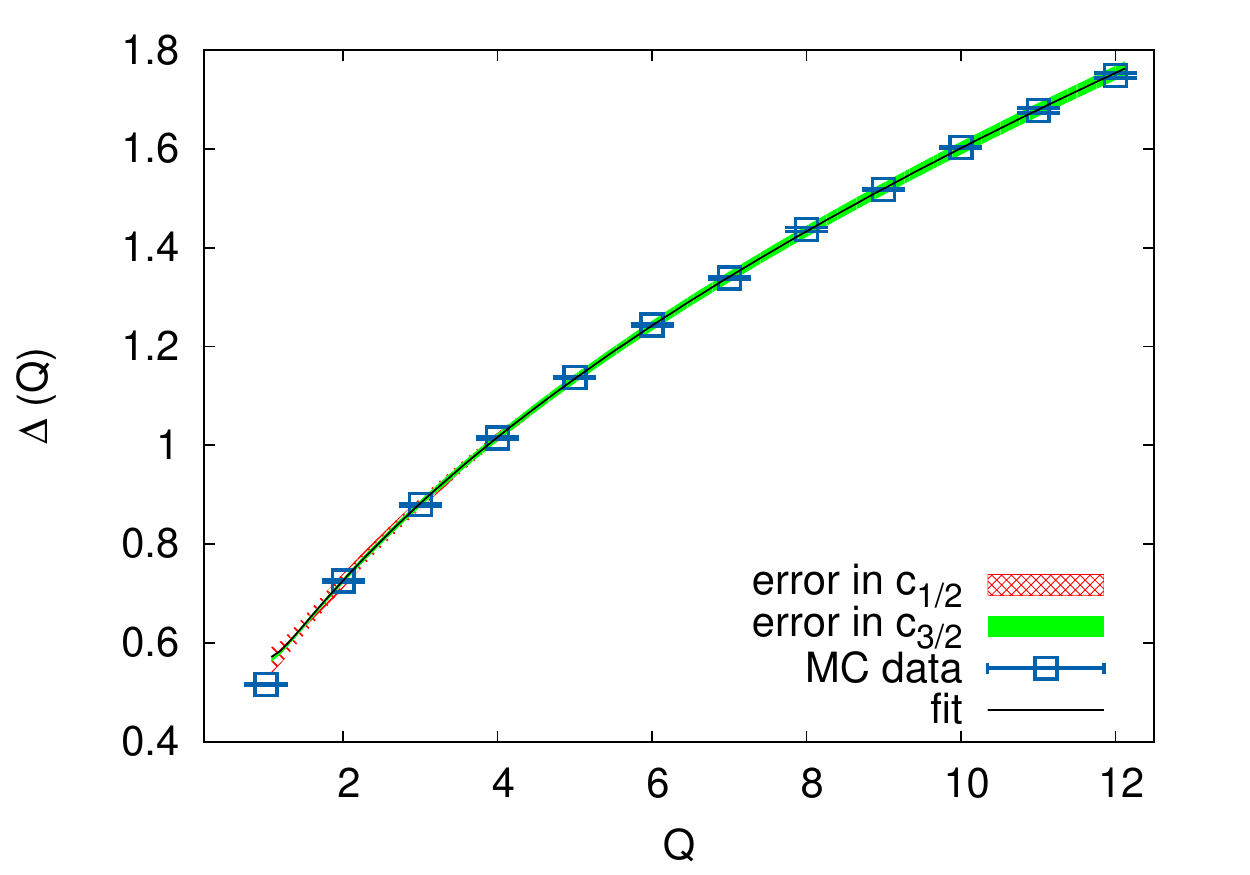}
 \caption{\label{fig:dqfits} Plot of $\Delta (Q)$ extracted using Monte Carlo calculations, along with a plot of Eq.~\eqref{eq:conformal-dimensions} keeping the first three non-zero coefficients estimated using our fits.}
 \end{figure}

Next we explore if we can connect our above calculations of $c_{\frac{3}{2}}$ and $c_{\frac{1}{2}}$ with the ones appearing in Eq.~(\eqref{eq:classical-energy}) for the expansion of the energy on a torus. Lattice calculations naturally lead to a torus geometry in the continuum limit if we keep the physical length $L$ fixed while taking the number of lattice points in each direction, $L/a$, to infinity. The lattice spacing $a$ itself is defined by setting the lattice energy $E^L(Q)$ to be equal to the continuum energy $E_{T^2}(Q)$ on the torus of side $L$ as the continuum limit is taken. On the lattice we measure the energy in terms of the dimensionless number $E^L(Q) a_t$ as a function of $L/a$, where $a_t$ is the temporal lattice spacing. Although the lattice calculation at a fixed $L/a$ and charge $Q$ breaks the symmetry between space and time, in the continuum limit (i.e., $L/a \rightarrow \infty$), we expect $a_t/a \rightarrow 1$ due to the cubic symmetry of the lattice action, Eq.~\eqref{eq:act}.  Thus on the lattice with a fixed $L/a$ we can replace Eq.~\eqref{eq:classical-energy} by the formula,
\begin{equation}
  \label{eq:lattice-energy}
  E^L(Q) L =  Q^{\frac{3}{2}}\Bigg(c^L_{\frac{3}{2}}+ c^L_{\frac{1}{2}} \ Q^{-1} + c^L_{-1/2}\ Q^{-2} \dots \Bigg) + q^L,
\end{equation}
such that in the continuum limit (i.e., large lattices) we expect this equation to turn into Eq.~\eqref{eq:classical-energy} on the torus, with $E^L(Q) \rightarrow E_{T^2}(Q)$, $c^L_{\frac{3}{2}} \rightarrow c_{\frac{3}{2}}$, $c^L_{\frac{1}{2}} \rightarrow (RL^2/2)c_{\frac{1}{2}} = 0$ and $q^L \rightarrow q_{T^2} = -0.508$.

Unfortunately, computing $E^L(Q)$ on the lattice is subtle due to the additive renormalization of lattice energies \cite{PhysRevE.69.046119,Hasenbusch:2008zz,HasenbuschP10006}. Hence in this work we use the techniques discussed in~\cite{PhysRevD.81.125007} to compute energy differences $\Delta E^L(Q) \ = \ (E^L(Q) - E^L(Q-1))$ at a fixed lattice size $L/a$. The idea is to couple a chemical potential $\mu a_t$ to the conserved charge $Q$ and extract the energy differences between ground states with charges $Q$ and $Q-1$ by tuning the chemical  potential. At the critical chemical potential $\mu_c^{(Q-1)} = \Delta E^L(Q) $ the average charge of the system jumps from $Q-1$ to $Q$. Further details on the method can be found in the supplementary material.

Unfortunately, our method is efficient only on small lattice sizes $L/a$, limiting the range of $Q$'s that can be used. Remember that the EFT description is valid only when $1 \ll (L/a)/\sqrt{Q} \ll L/a$. Further, small lattices also imply larger lattice spacing errors which means $c^L_{\frac{3}{2}}$ and $c^L_{\frac{1}{2}}$ may not quite agree with continuum expectations discussed above. Still we can learn about the magnitude of the errors. With this in mind we have computed $\Delta E^L(Q) L$ in the range $1 \leq Q \leq 18$ for $L/a=8$ and $10$. Our results are tabulated in Table~\ref{tab:lattice-energy}.
\begin{table}[!htb]
  \begin{tabular}{ccc||ccc}
    \toprule
    \(Q\)  & \multicolumn{2}{c||}{$\Delta E^L(Q) L$ } &
    \(Q\)  & \multicolumn{2}{c}{$\Delta E^L(Q) L$ } \\
       & $(L/a=8)$          & $(L/a = 10)$       &
       & $(L/a=8)$          & $(L/a = 10)$       \\
    \colrule
    1  & 1.3442(5)          & 1.3393(7)          &
    10 & 5.7135(24)         & 5.6998(3)          \\
    2  & 2.2422(3)          & 2.2311(4)          &
    11 & 6.0074(24)         & 5.9960(4)          \\
    3  & 2.9012(4)          & 2.8851(3)          &
    12 & 6.2866(24)         & 6.2786(4)          \\
    4  & 3.4434(3)          & 3.4259(3)          &
    13 & 6.5529(24)         & 6.5487(4)          \\
    5  & 3.9142(3)          & 3.8949(2)          &
    14 & 6.8078(32)         & 6.8074(5)          \\
    6  & 4.3346(16)         & 4.3150(2)          &
    15 & 7.0524(32)         & 7.0560(20)         \\
    7  & 4.7178(16)         & 4.6987(2)          &
    16 & 7.2884(32)         & 7.2970(10)         \\
    8  & 5.0722(16)         & 5.0545(3)          &
    17 & 7.5152(32)         & 7.5280(20)         \\
    9  & 5.4026(16)         & 5.3868(3)          &
    18 & -                  & 7.7530(20)          \\
    \botrule
  \end{tabular}
  \caption{Values of $\Delta E^L(Q) L$ obtained on the lattice using the Monte Carlo method for various values of $Q$ on lattice sizes $L/a=8$ and $10$. \label{tab:lattice-energy}}
\end{table}

\begin{figure}[!htb]
  \includegraphics[width=0.45\textwidth]{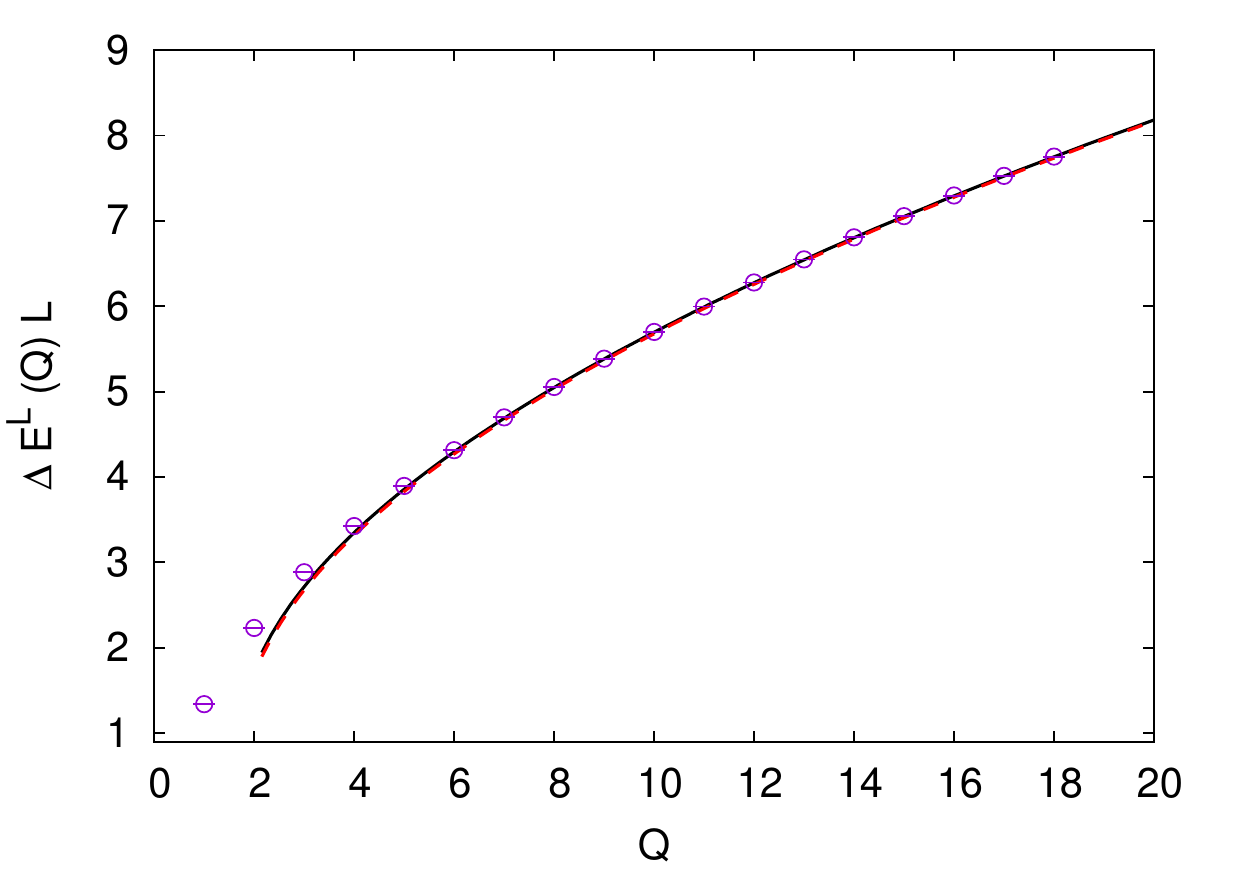}
  \caption{\label{fig:enplot} The plot of $\Delta E^L(Q) L$ for $L/a = 10$ given in Table~\ref{tab:lattice-energy} as a function of $Q$ along with a plot of the fitted function (solid line) with the fit parameters $c^L_{\frac{3}{2}}=1.235$ and $c^L_{\frac{1}{2}}=0.12$ and $c^L_{-\frac{1}{2}}=1.9$. The function with $c^L_{\frac{1}{2}} = 0$ (dashed line) almost coincides with the solid line.}
\end{figure}

Our data for $L/a=8$ and $10$ fits well to Eq.~\eqref{eq:lattice-energy} with a $\chi^2/DOF \approx 1$, as long as we restrict the fits to the range $5 \leq Q \leq 13$ and $10 \leq Q \leq 18$ respectively. The fit result for $L/a=10$ is shown in Fig.~\ref{fig:enplot}. The fits give $c^L_{\frac{3}{2}} = 1.235(10)$ and $c^L_{\frac{1}{2}} = 0.12(10)$ for both sets of the data. However, $c^L_{-\frac{1}{2}}$ fits to $1.9(5)$ at $L/a=10$ and to $0.7(2)$ at $L/a=8$. Note that $c^L_{\frac{3}{2}}$ is only about $3$\% off from $1.195(10)$ extracted earlier using conformal dimensions. This error seems reasonable given our small lattices (see the supplementary material for further discussion of this point). The coefficient $c^L_{\frac{1}{2}}$ is small, perhaps related to the fact that it is expected to be zero in the continuum. The value of $c_{-\frac{1}{2}}$ is consistent with being non-zero, although it was recently argued that the classical contribution to the energy at this order is expected to vanish in the continuum limit~\cite{Loukas:2016ckj}. Contributions from quantum corrections were not computed and could in principle be non-zero. Finally, as discussed in the supplementary material, all of our results are consistent with those obtained recently \cite{Whitsitt:2017ocl}.

We wish to thank Uwe-Jens~Wiese for helpful conversations. D.O. would like to thank Antonio~Amariti, Simeon~Hellerman, Orestis~Loukas and Susanne~Reffert for enlightening discussions and comments. D.B. would like to thank Martin~Hasenbusch, Ferenc~Niedermayer, Rainer~Sommer and~Ulli Wolff for useful discussions. The material presented here is based upon work supported by the U.S. Department of Energy, Office of Science, Nuclear Physics program under Award Number DE-FG02-05ER41368.

\bibliography{LargeQ}
\newpage

 \section{Supplementary Material: the algorithm}
 As explained in \cite{PhysRevD.81.125007}, we can construct a worm algorithm to compute quantities in statistical mechanics, starting with the classical action
\begin{equation}
  S([\theta]) = -\beta \sum_{\mathrm{x},\alpha} \cos( \theta_{\mathrm{x}} - \theta_{\mathrm{x}+\hat{\alpha} a}).
\end{equation}
We first use the identity
\begin{equation}
\exp{ \beta \cos \theta} = \sum_{k = -\infty}^{\infty} I_k (\beta) e^{i k \theta},
 \end{equation}
on each bond $(x,\alpha)$ to integrate out the field variables ($\theta_{x}$) from the partition function
 \begin{equation}
 Z \ =\ \int [ d\theta_x] \ e^{-S([\theta])},
 \end{equation}
and express it in terms of a configuration of integer bond variables $k_{x,\alpha}$, each of which is an integer valued worldline variable denoting the charge flowing on the bond $(x,\alpha)$, from $x$ to $(x+\hat{\alpha})$. The function $I_k(\beta)$ is the modified Bessel function of the first kind. In terms of worldline configurations $[k]$, the partition function looks like
\begin{equation}
 Z = \sum_{[k]} \prod_{x,\alpha} \left\{ I_{k_{x,\alpha}} (\beta) \right\} \ \prod_x\ \delta \left( \sum_\alpha (k_{x,\alpha} - k_{x-\hat{\alpha},\alpha}) \right).
 \end{equation}
We can use the standard worm algorithm to update worldline configurations $[k]$ as follows:
\begin{enumerate}
\item We pick a random site on the lattice, $x = x_h$ and refer to it as the head site. At the beginning we also define the tail site to be the same as the head site, $x_t = x_h$.
\item We randomly pick one of the $2d$ neighbors $x_t + \hat{\alpha}, \alpha = \pm 1, \pm 2, \cdots, \pm d$.
\item Let $k$ be the current on the bond joining $x_t$ and $x_t+\hat{\alpha}$. If $\alpha > 0$, we update the forward current from $k$ to $k+1$ with probability $I_{k+1}(\beta)/I_k(\beta)$. If this update is accepted, then we change the tail site from $x_t \rightarrow x_t+\hat{\alpha}$. If $\alpha < 0$, then with probability $I_{k-1}(\beta)/I_k(\beta)$, we update $k$ to $k-1$ and set $x_t \rightarrow x_t+\hat{\alpha}$. When the updates are not accepted, $x_t$ remains the same.
\item If the tail site $x_t$ reaches the head site $x_h$ the worm update ends. Otherwise, we go back to step 2.
\end{enumerate}
In the above update, when we start we introduce a single positive charge at $x_h$ and a single negative charge at $x_t$. During the worm update the tail site $x_t$ moves around the lattice until it comes back and annihilates with the head site. However, at each of the intermediate steps the worm update samples the two-point correlation function $C_1(x_t,x_h)  = \langle e^{i \theta_{x_h}} e^{-i \theta_{x_t}}\rangle$.

 \begin{figure}[!htb]
 \includegraphics[width=0.2\textwidth]{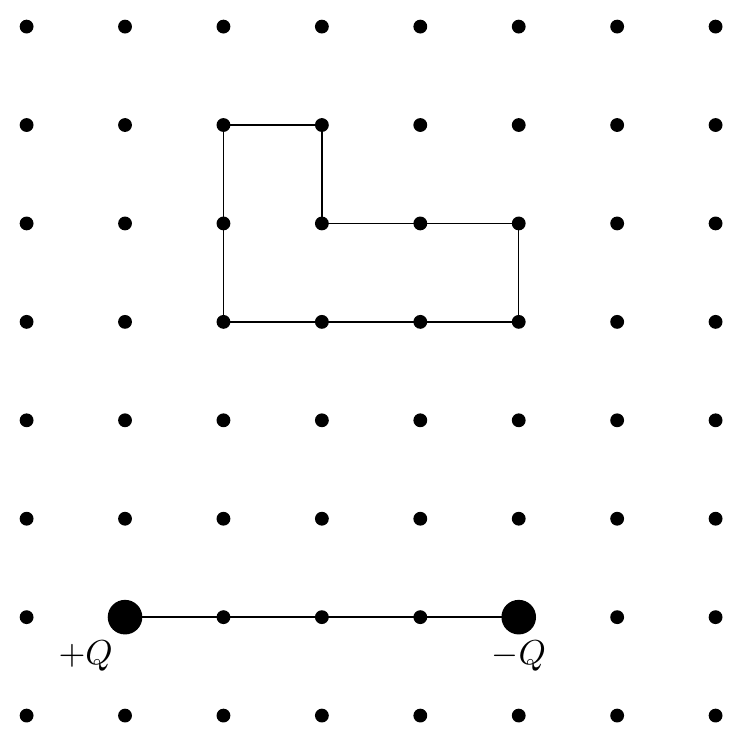}
 \hspace{1cm}
 \includegraphics[width=0.2\textwidth]{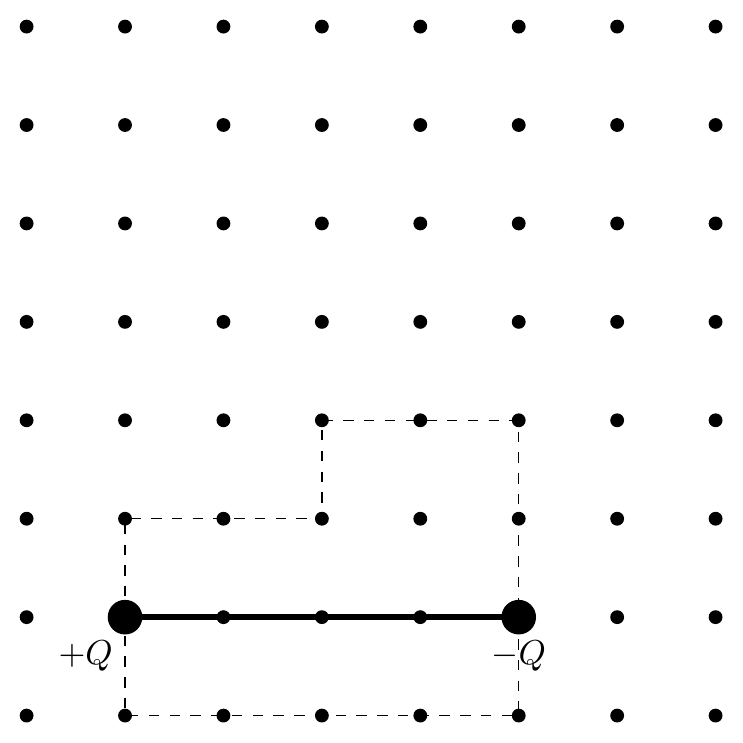}
 \caption{\label{fig:worm} (left) A standard worm update which updates the background together with a source $Q$ and sink $-Q$ at a fixed distance apart, (right) an example measurement update.}
 \end{figure}

We can easily adapt the worm algorithm to measure the two point correlation function $C_{Q}(r) = \langle e^{i Q \theta_0} e^{-i Q\theta_r}\rangle$. We simply introduce charge $Q$ at the head site and a charge $-Q$ at the tail site by updating the bonds by changing $k$ to $k\pm Q$ as we move. Whenever the tail reaches the site $x_t = x_h+r$, we get a contribution to $C_Q(r)$. Combining this with the standard worm algorithm we can in principle get an ergodic algorithm. However, such an algorithm leads to a severe signal-to-noise problem, since the updates of $k$ to $k\pm Q$ is extremely inefficient especially for large $Q$. To solve this problem we focus on the ratio of correlation functions
\begin{equation}
R(r) = \frac{C_{Q+1} (r)}{C_Q(r)} = \frac{Z_{Q+1,r}}{Z_{Q,r}},
\end{equation}
where we define
\begin{multline}
 Z_{Q,r} = \sum_{[k]} \prod_{x,\alpha} \left\{ I_{k_{x,\alpha}} (\beta) \right\} \
 \\
 \prod_x\ \delta 
\left( \sum_\alpha (k_{x,\alpha} - k_{x-\hat{\alpha},\alpha}) - Q_{x}\right), 
\end{multline}
with $Q_x=Q (\delta_{x,o} - \delta_{x,r})$. In other words $Z_{Q,r}$ is a partition function with a charge $Q$ fixed at $x=0$ and $-Q$ at $x=r$. Note that we can also write the above ratio slightly differently as
 \begin{equation}
 R(r) =\langle e^{i \theta_{r}} e^{-i \theta_0}\rangle_{Q,r},
 \end{equation}
where the expectation value on the right is now taken in the distribution of $[k]$ according to the partition function $Z_{Q,r}$. In this case the standard worm algorithm can be used to update the configurations $[k]$ associated with $Z_{Q,r}$, which we refer to as the target ensemble. In Fig.~\ref{fig:worm} (left) we show an illustration of a configuration in the target ensemble with a worm loop that updates the background configuration. In addition to the standard algorithm, we construct a special measurement algorithm where the worm update starts and ends at $x_h=0$. If during this measurement update the tail site touches reaches the site $x_t=r$ we count it as a contribution to the ratio $R(r)$. More concretely,
\begin{enumerate}
 \item We start with a configuration which has $Q$ charges at site $(0,0,0)$ and $-Q$ charge at site $(r,0,0)$, and initialize a counter $c=0$.
 \item We begin a worm update with the head at $x_h = (0,0,0)$. We also define the tail site as $x_t=x_h$.
 \item We pick one of the $2d$ neighbors of $x_t$, and propagate the tail site using the standard worm update described earlier.
 \item Whenever $x_t = (r,0,0)$, we increment the counter $c = c+1$. This implies that the configuration generated contributes to the ratio $R(r)$.
 \item When the tail site returns to the head site the update ends.
\end{enumerate}
An illustration of this measurement update is shown in Fig.~\ref{fig:worm} (right), where the worm starts from $x_h=0$ and passes through $x_t=r$ before closing. $R(r)$ is computed as an average of the value of $c$ measured for each measurement worm algorithm. The actual algorithm involves several standard worm updates and an equal number of measurement updates.

For our simulations, we work at several finite cubic volumes with linear size $L$ and compute the correlation ratios with $r=L/2$. In this calculation, the flux from charge $Q$ can reach the sink either directly through the bulk or through the boundary. We can define a winding number $w$ for each configuration $[k]$ as the number of flux lines that reach the sink through the boundary. Then the remaining $Q-w$ flux lines reach through the bulk. We have discovered that the value of $R(r)$ is sensitive to $w$. In our work we sample all possible windings distributed according to $Z_{Q ,r}$. On an average we see that $w=Q/2$. The standard worm algorithm is able to change the winding number $w$ quite efficiently.

 \section{Details of fitting and error analysis}
 In the main text, we have presented evidence for the validity of the large charge analysis developed in~\cite{Hellerman:2015nra,Alvarez-Gaume:2016vff}. In this Supplementary we first explain the analysis details to compare the theory with the Monte Carlo 
data, as well as model (in)-dependence of our results. We also compare our results with some related results in the literature~\cite{Whitsitt:2017ocl} for the U(1) global charge in $(2+1)-$dimensions.

\subsection{Conformal Dimensions}
  The main text, and the first supplementary section describes rather extensively the new method developed to estimate the difference in the conformal dimensions, $D(Q) - D(Q-1)$ directly from the Monte Carlo simulations. Absolute conformal dimension $D(Q)$ are obtained by using (with $D(0) = 0$):
\begin{align}
 \nonumber
 D(Q) &=  [D(Q) - D(Q-1)] + [D(Q-1) - D(Q-2)] \\
      &+ \cdots + [D(1) - D(0)].
\end{align}
    Since the individual terms, $D(Q) - D(Q-1)$, are obtained by fits to completely different sets of independent simulations, there is no correlation between the different sets and the errors can be added in quadrature to obtain the error on $D(Q)$:
\begin{align}
 \nonumber
 \delta[D(Q)]^2 &= \delta[D(Q) - D(Q-1)]^2 \\
 \nonumber      &+ \delta[D(Q-1) - D(Q-2)]^2 \\
                &+ \cdots +\delta[D(1) - D(0)]^2.  
\end{align}
 A bootstrap analysis was also done to check the error-bars on these values.

\emph{Fit methodology and model (in) dependence:} As emphasized in the main text, we have tried to verify the validity of the effective theory developed in~\cite{Hellerman:2015nra,Alvarez-Gaume:2016vff}, which predicts the conformal dimensions $D(Q)$ as in Eq. (1). However, from the Monte Carlo results we can directly measure only the difference $D(Q) - D(Q-1)$. Together with the input of $D(Q=0) = 0$, we can reconstruct the values for $D(Q)$. To verify the effective theory (EFT), we fixed the value of $c_0$ from the EFT and then fitted both the difference of the conformal dimensions, and the total conformal dimensions to their respective functional forms (Eq. (8)) to extract the low energy coefficients $c_\frac{3}{2}$ and $c_\frac{1}{2}$ simultaneously. The results of this fit are described in the main text.

  We note that the systematic power law expansion of the conformal dimensions in $1/Q$ is rather robust, it is possible to relax the input from the EFT for the value of the additive constant $c_0$, and determine it completely from the numerical results by treating it as a fit parameter. For example, keeping the constants $c_{\frac{3}{2}}$ and $c_{\frac{1}{2}}$ fixed to their values as described in the methodology before, we have tried to fit $c_0$ from the $D(Q)$ results. In this case, we obtained for $c_0$ the value -0.102(4), where the errorbar takes into account the different fit ranges, about a 2-sigma difference from the value predicted by the EFT. We also tried the different strategy of fitting all the coefficients $c_{\frac{3}{2}}$, $c_{\frac{1}{2}}$ and $c_0$ from the data on $D(Q)$. This yields: $c_{\frac{3}{2}} = 1.195(5)$, $c_{\frac{1}{2}} = 0.07(1)$ and $c_0 = -0.09(2)$. As one notes, these values are completely consistent with the results quoted in the main text, as well the input from the EFT. We note that these different strategies show that the assumed functional form for the conformal dimensions is rather robust, and gives confidence in the model independence of the results. Another important point in the analysis is that from the ansatz of Eq.~(1), there is a contribution from the term $c_{-\frac{1}{2}} (4 \pi/Q)^2$, which is however very suppressed, even with the accuracy of our data. This contaminates the extraction of either the term $c_0$, or the estimate of $c_{-\frac{1}{2}}$. In testing the different strategies, we have favored ranges with larger starting values, such that the latter term can be fixed to zero in the fit.

\subsection{Energy} 
  In this subsection, we explain the methodology for the energy computation, used in ~\cite{PhysRevD.81.125007}. This is completely different method used for calculating the conformal dimensions and provide an independent way of checking the EFT. The basic idea is to couple a chemical potential $\mu$ to the conserved global $U(1)$ charge Q. As the chemical potential is increased, the ground state is the one which has a total charge $Q$, since the chemical potential couples as $H - \mu Q$. The action for the XY-model in the presence of the
chemical potential is
\begin{equation}
S([\theta])=-\beta \sum_{\mathrm{x},\alpha} \cos( \theta_{\mathrm{x}} - \theta_{\mathrm{x}+\hat{\alpha} a} - i \mu \delta_{\alpha,t}).
\end{equation}
 Thanks to the worm algorithm, this model can be simulated with a chemical potential as demonstrated in ~\cite{PhysRevD.81.125007}. It was verified that with increase in chemical potential there are level crossing phenomena between a state $E^{(Q-1)}$ with a charge $Q-1$ and a state $E^Q$. With increasing charge, the latter becomes the ground state of the system. Close to the critical chemical potential where the level crossing occurs, one can approximate the partition function as
\begin{equation}
 Z \approx \mathrm{e}^{-(E_0^{(Q-1)} - \mu (Q-1)) L_t} + \mathrm{e}^{-(E_0^{(Q)} - \mu Q) L_t},  
 \label{eq:partF}
\end{equation}
 where $E_0^{Q}$ is the ground state in the charge $Q$ sector. We have assumed that all the higher energy states are suppressed exponentially at large $L_t$. It is easy to verify that $\mu_c^{(Q-1)} =\ (E_0^{(Q)} - E_0^{(Q-1)})$ is the critical  chemical potential where the state with charge $Q$ becomes the ground state instead of the state with charge $Q-1$. Since these calculations are performed at a fixed lattice size $L$ we define $\Delta E^L(Q) = \mu_c^{(Q-1)}$ and these are the numbers reported in Table 3 of the main text. 

\begin{figure}
\includegraphics[scale=0.7]{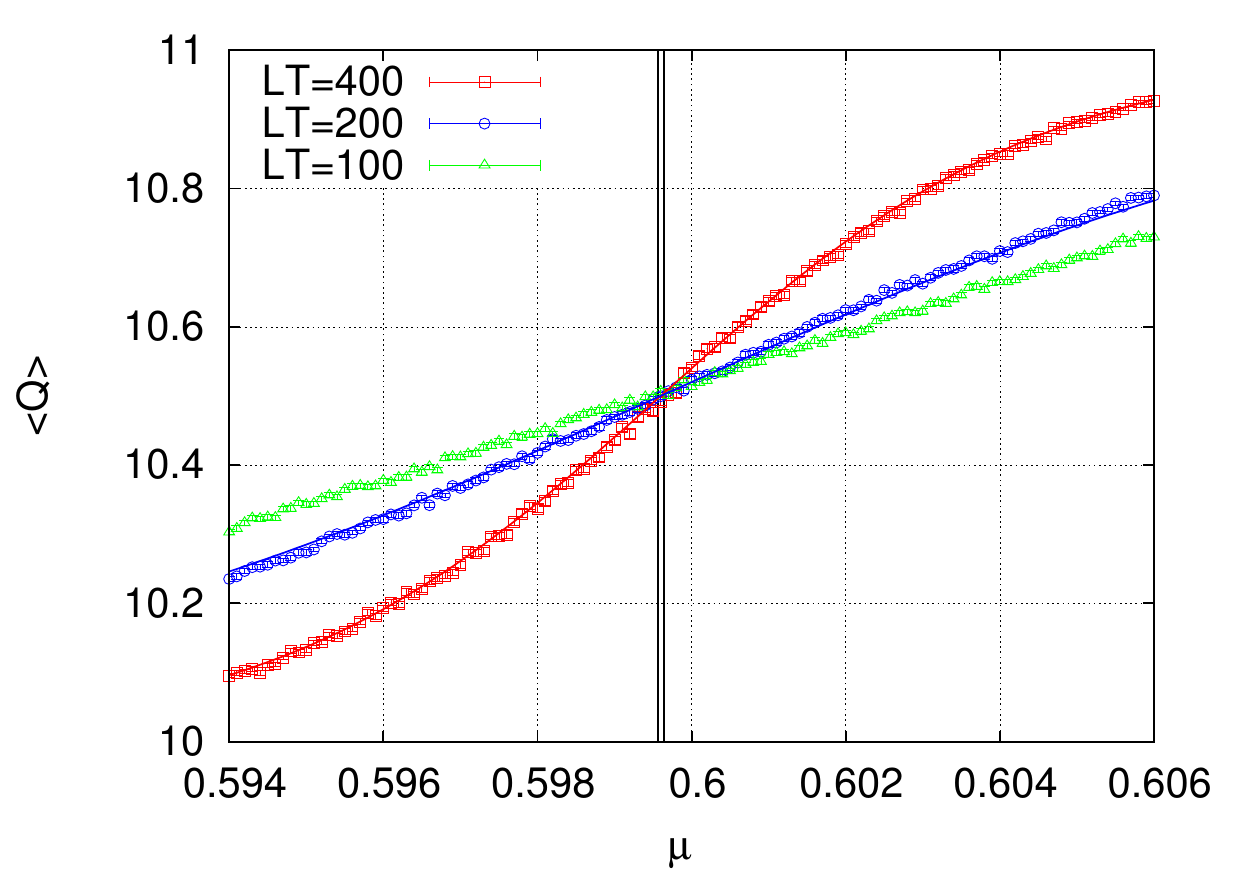}
\caption{The extraction of the critical chemical potential $\mu_c^{(10)}$ by fitting the Eq. \ref{eq:fitEN} to the level crossing between the states $Q=10$ and $Q=11$. The curves passing through the red squares and the blue circles show the combined fit of the data to a single parameter fit, $\mu_c^{(10)}$. The fit does not fit to the data for $L_t=100$, and hence one needs larger values of $L_t$. The vertical 1-sigma band shows the error on the extracted critical chemical potential.}
\label{fitEN}
\end{figure}

 The computation of $\mu_c^{(Q)}$ is done by measuring the average charge density as a function of the chemical potential, which is given as
\begin{equation}
 \frac{\langle Q \rangle}{L^2} = \frac{Q + (Q+1) \mathrm{e}^{\Delta^{(Q)}_\mu L_t}}{1 + \mathrm{e}^{\Delta^{(Q)}_\mu L_t}} 
 \label{eq:fitEN}
\end{equation}
 where $\Delta_\mu^{(Q)} = \mu - \mu_c^{(Q)}$. The use of large $L_t$ is essential to suppress the higher excitations such that Eq.~(\ref{eq:fitEN}) provides a reliable fit to the data. We have done further computations to confirm our error analysis. As an example, in Fig \ref{fitEN} we show the data we used to fit to  Eq.~(\ref{eq:fitEN}) and extract $\mu_c^{(Q)}$ and its error from it.

\begin{figure*}
\includegraphics[scale=0.7]{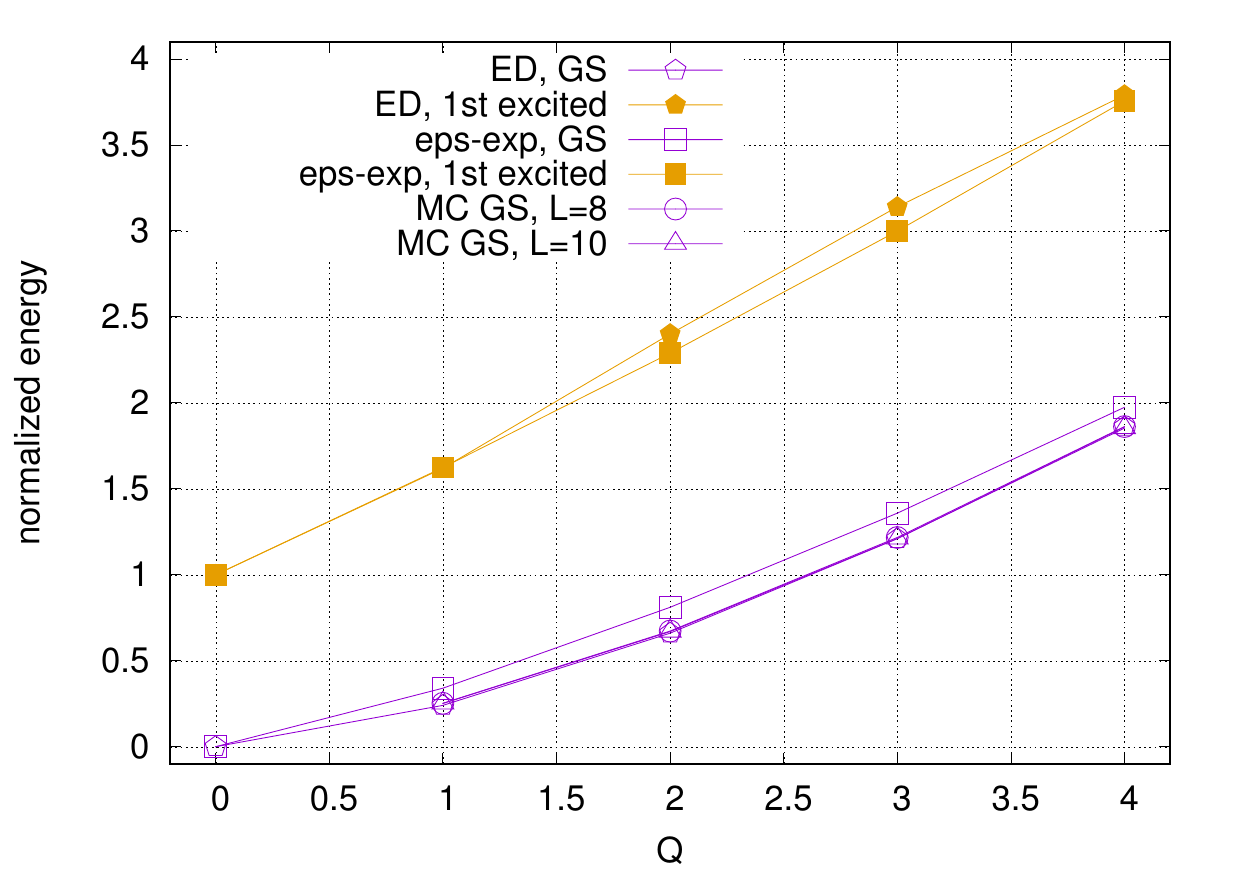}
\caption{Ground state (GS) and first excited state energies in the charge sectors $Q=0,1,2,3,4$ taken from Table 2 of Ref. \cite{Whitsitt:2017ocl}. Following Ref.\cite{Whitsitt:2017ocl}, the ground state at $Q=0$ is taken to be zero and the first excited state at $Q=0$ is normalized to unity. The open and filled pentagons show results from
exact diagonalization studies (ED) for the ground state (GS) and first excited (1st excited) states respectively. Open and filled squares show the same results obtained from epsilon expansion (eps-exp) studies. The circles and triangles are our Monte Carlo (MC) results for the ground state at $L=8$ and $10$ respectively and are in agreement with previous work.}
\label{enL810}
\end{figure*}

\subsection{Corrections to scaling}
 In any scaling analysis of numerical data, corrections to scaling can only be addressed when the leading data does not match the predicted scaling form. In the case when the leading scaling form is obeyed by the data, corrections to the scaling are assumed to be negligible. Otherwise, the fitting becomes an ill-defined problem. Our data for the ratios of the correlation functions, $R(L/2)$ in Eq. (7), are consistent with the leading order scaling form, as shown in the Fig 2 (top and bottom). This suggests that, at the level of accuracy of our data (at the per mille level), the corrections to scaling are negligible.

It is important to note that our results for the conformal dimensions is extracted from much bigger lattice sizes and for a large range of lattice sizes than our results for the Energy.  Note that the lattices which are used to extract the conformal dimensions ($L/a > 48$) are much bigger than $L=8,10$ used in the energy computation.  Thus, we do believe that our energy results do contain corrections to scaling. In fact the coefficient $c_{\frac{3}{2}} = 1.235(10)$ extracted from the energy computation differs from the coefficient $c_{\frac{3}{2}} = 1.195(10)$ by 3\%, which we believe is indeed due to scaling violations (or equivalently lattice spacing errors). According to studies in literature \cite{Hasenbusch:1999cc,Campostrini:2006ms}, the leading correction to scaling at the $O(2)$ fixed point is $\omega \approx 0.8$. Using the relation
\begin{equation}
 c_{\frac{3}{2}}^{L=\infty} \approx c_{\frac{3}{2}}^{L} \left( 1 + \frac{\gamma}{L^\omega} \right)
 \label{eq:scalingcorr} 
\end{equation}
and substituting $c_{\frac{3}{2}}^{L=\infty} = 1.195$ and $c_{\frac{3}{2}}^{L=10}=1.135$ and $\omega = 0.8$, we obtain $\gamma \sim -0.4$, which seems like a reasonable estimate of the scaling correction term. Further, also note that in the ratio of  correlation functions, the leading order scaling corrections get canceled:
\begin{align}
\nonumber R(L/2) &\sim \frac{1}{L^{2 D(Q)}}\left( 1 + \frac{d}{L^\omega} \right) 
    \left[\frac{1}{L^{2 D(Q-1)}}\left( 1 + \frac{d}{L^\omega} \right) \right]^{-1} \\ 
   &\sim \frac{1}{L^{2 D(Q)- 2 D(Q-1)}} \left( 1 - \frac{d^2}{L^{2 \omega}} \right).
\end{align}  
For lattices $L/a > 48$ this correction is less than the statistical errors reported in Table 2 of the main text, and explains why the scaling corrections were not essential in the extraction of conformal dimensions.

\subsection{Comparison with Results of ~\cite{Whitsitt:2017ocl} }
 Recently, there has been some interesting work (see~\cite{PhysRevLett.117.210401,Whitsitt:2017ocl}), in which the authors compute the energy spectrum of an $O(2)$ symmetric Hamiltonian for different global charge sectors on a finite torus, using both the analytic epsilon-expansion (eps-exp) technique, as well as numerical calculations using exact diagonalization methods (ED).
The results are given in Table 2 of Ref. ~\cite{Whitsitt:2017ocl}. It is important to note that in the ED method it is difficult to compute the exact ground state energy as a function of $Q$, and hence the energy at $Q=0$ was set to zero. Also, energy units were chosen such that the energy gap to the first excited state in the $Q=0$ sector is 
normalized to unity. With these choices the spectrum computed in ~\cite{Whitsitt:2017ocl} in the two methods is plotted in Fig \ref{enL810}.  It is very useful to compare our results for the ground state with these results. If we normalize our results similarly, our values for the ground state energies for both L=8,10 lattices, show good agreement with the work of ~\cite{Whitsitt:2017ocl} as shown in Fig \ref{enL810}.

\end{document}